\documentclass[prd,article,twocolumn,preprintnumbers,superscriptaddress,amsmath,amssymb,aps,longbibliography]{revtex4-1}
\usepackage{graphicx}
\usepackage{amssymb}
\usepackage{natbib}
\usepackage{hyperref}
\usepackage{mathtools}
\usepackage{slashed}
\usepackage{url}
\usepackage{graphicx}
\usepackage{cleveref}
\usepackage[compat=1.0.0]{tikz-feynman}
\usepackage{tikz-feynman}
\usepackage{overpic}
\usepackage{subfigure}

\usepackage{siunitx}

\DeclareSIUnit \s {\second}
\DeclareSIUnit \ns {\nano\second}
\DeclareSIUnit \mus {\micro\second}
\DeclareSIUnit \ms {\milli\second}
\DeclareSIUnit \MB {\mega\byte}
\DeclareSIUnit \GB {\giga\byte}
\DeclareSIUnit \TB {\tera\byte}
\DeclareSIUnit \PB {\peta\byte}
\DeclareSIUnit \Mbps {\mega\bit/\s}
\DeclareSIUnit \Gbps {\giga\bit/\s}
\DeclareSIUnit \Tbps {\tera\bit/\s}
\DeclareSIUnit \Pbps {\peta\bit/\s}
\DeclareSIUnit \kton {\kilo\tonne} 
\DeclareSIUnit \kt {\kilo\tonne}
\DeclareSIUnit \Mt {\mega\tonne}
\DeclareSIUnit \eV {\electronvolt}
\DeclareSIUnit \keV {\kilo\electronvolt}
\DeclareSIUnit \MeV {\mega\electronvolt}
\DeclareSIUnit \GeV {\giga\electronvolt}
\DeclareSIUnit \TeV {\tera\electronvolt}
\DeclareSIUnit \PeV {\peta\electronvolt}
\DeclareSIUnit \EeV {\exa\electronvolt}
\DeclareSIUnit \m {\meter}
\DeclareSIUnit \cm {\centi\meter}
\DeclareSIUnit \in {\inchcommand}
\DeclareSIUnit \km {\kilo\meter}
\DeclareSIUnit \kV {\kilo\volt}
\DeclareSIUnit \kW {\kilo\watt}
\DeclareSIUnit \MW {\mega\watt}
\DeclareSIUnit \MHz {\mega\hertz}
\DeclareSIUnit \mrad {\milli\radian}
\DeclareSIUnit \year {years}
\DeclareSIUnit \POT {POT}
\DeclareSIUnit \sig {$\sigma$}
\DeclareSIUnit\parsec{pc}
\DeclareSIUnit\lightyear{ly}
\DeclareSIUnit\foot{ft}
\DeclareSIUnit\ft{ft}
\DeclareSIUnit \ppb{ppb}
\DeclareSIUnit \ppt{ppt}
\DeclareSIUnit \samples{S}
\DeclareSIUnit \pe{PE}
\DeclareSIUnit \T{T}

\newcommand{\enu}{\E_\enu}

\def\gs{\mathrel{
   \rlap{\raise 0.511ex \hbox{$>$}}{\lower 0.511ex \hbox{$\sim$}}}}
\def\ls{\mathrel{
   \rlap{\raise 0.511ex \hbox{$<$}}{\lower 0.511ex \hbox{$\sim$}}}} 
\newcommand{\thot}{\frac{\theta}{2}}

\begin{document}
\title{Tau depolarization at very high energies for neutrino telescopes}

\author{Carlos A. Arg\"uelles}
\affiliation{Department of Physics \& Laboratory for Particle Physics and Cosmology, Harvard University, Cambridge, MA 02138, USA}
\author{Diksha Garg}
\affiliation{Department of Physics and Astronomy, University of Iowa,
Iowa City, Iowa 52242, USA}
\author{Sameer Patel}
\affiliation{Department of Physics and Astronomy, University of Iowa,
Iowa City, Iowa 52242, USA}
\author{Mary Hall Reno}
\affiliation{Department of Physics and Astronomy, University of Iowa,
Iowa City, Iowa 52242, USA}
\author{Ibrahim Safa}
\affiliation{Department of Physics \& Laboratory for Particle Physics and Cosmology, Harvard University, Cambridge, MA 02138, USA}
\affiliation{Department of Physics \& Wisconsin IceCube Particle Astrophysics Center, University of Wisconsin-Madison, Madison, WI 53706, USA}
\date{\today}

\begin{abstract}
    The neutrino interaction length scales with energy, and becomes comparable to Earth’s diameter above 10's of TeV energies. Over terrestrial distances, the tau's short lifetime leads to an energetic regenerated tau neutrino flux, $\nu_\tau\to\tau\to \nu_\tau$, within the Earth.
    The next generation of neutrino experiments aim to detect ultra-high energy neutrinos. Many of them rely on detecting either the regenerated tau neutrino, or a tau decay shower.
    Both of these signatures can be affected by the polarization of the tau through the energy distribution of the secondary particles produced from the tau's decay. While taus produced in weak interactions are nearly 100\% polarized, it is expected that taus experience some depolarization due to electromagnetic interactions in the Earth.
    In this paper, for the first time we quantify the depolarization of taus in electromagnetic energy loss.
    We find that tau depolarization has only small effects on the final energy of tau neutrinos or taus produced by high energy tau neutrinos incident on the Earth. Tau depolarization can be directly implemented in Monte Carlo simulations such as \texttt{nuPyProp} and \texttt{TauRunner}. 
\end{abstract}
\maketitle

\tableofcontents

\section{Introduction}

The detection of solar and atmospheric muon- and electron-neutrinos through their interactions in large underground detectors has led to our current understanding of neutrino masses and oscillations~\cite{Fukuda:1998mi,Ahmad:2001an,Ahmad:2002jz}.
Over distance scales characterized by the diameter of the Earth, for energies in the range of ${\cal O} (1)$--${\cal O}$(10) GeV, the disappearance of muon neutrinos from oscillations~\cite{Super-Kamiokande:2017yvm} and the corresponding tau neutrino appearance~\cite{Super-Kamiokande:2017yvm,Super-Kamiokande:2017edb,IceCube:2019dqi} at Super-Kamiokande and IceCube-DeepCore highlight the role of neutrino telescopes.
The first detection of a diffuse astrophysical neutrino flux by the IceCube Neutrino Observatory~\cite{Aartsen:2013jdh} established the field of high-energy neutrino astronomy.
Astrophysical neutrinos come from sources in which high-energy protons interact with ambient protons or photons to produce pions and other hadrons~\cite{Gaisser:1994yf,Learned:2000sw,Becker:2007sv}.
Pion decays, which are expected to dominate the flux, lead to $\nu_\mu$ and $\nu_e$ fluxes (and their anti-neutrino partners) in a ratio of approximately $2:1$ at the source.
Neutrino oscillations over astrophysical distances yield nearly equal fluxes of the three neutrino flavors~\cite{Learned:1994wg,Pakvasa:2007dc,Song:2020nfh}.
Measurements of neutrino flavor ratios hold the potential to characterize their sources, and may provide evidence of new physics scenarios at extreme energies (see, e.g., ref.~\cite{Beacom:2003nh,Bustamante:2010nq,Mehta:2011qb,Arguelles:2015dca,Brdar:2016thq,Rasmussen:2017ert,Arguelles:2019rbn,Bustamante:2019sdb,Farzan:2018pnk,Ahlers:2020miq,Bustamante:2018mzu}).
Through interactions of all three neutrino flavors in contained cascade events in IceCube and with through-going muons that originate from $\nu_\mu$ charged-current interactions, the diffuse neutrino flux has been measured up to neutrino energies in the PeV range~\cite{Schneider:2019ayi,Stettner:2019tok,IceCube:2020acn,IceCube:2020wum}.
IceCube's Glashow resonance event~\cite{IceCube:2021rpz} pushes the neutrino energy measurement to $E_\nu\simeq 6.3$~PeV.

In pursuit of neutrino probes of even higher energy phenomena, strategies for the detection of tau neutrinos have come to the fore~\cite{Abraham:2022jse}.
Within a detector like IceCube and its proposed successor IceCube-Gen2~\cite{IceCube-Gen2:2020qha}, double-pulse and so-called double-bang events with $\nu_\tau\to\tau$ production via charged-current interactions followed by $\tau$ decays will give distinct signals.
Already, two candidate $\nu_\tau$ events have been detected by IceCube~\cite{Abbasi:2020zmr}.
At higher energies, $\tau$'s produced outside the detector can appear as tracks with energetic decays.
For $E_\tau>100$ PeV, the tau track length before decay is $\gs 5$~km on average.
Thus, there is a measurable probability for very high energy (VHE) tau neutrinos to convert to $\tau$'s in the Earth which in turn emerge to produce air showers. 
Detection of these $\tau$-decay induced air showers are targets of current and future experiments such as ANITA~\cite{ANITA:2008mzi,ANITA:2020gmv}, 
PUEO~\cite{Deaconu:2019rdx,PUEO:2020bnn}, 
BEACON~\cite{Wissel:2020fav}, Trinity~\cite{Otte:2019aaf}, TAMBO~\cite{Romero-Wolf:2020pzh}, GRAND~\cite{Alvarez-Muniz:2018bhp}, 
EUSO-SPB2~\cite{Adams:2017fjh,Eser:2021mbp}, 
and POEMMA~\cite{Olinto:2020oky}. 

Detailed modeling of tau neutrino and tau propagation in Earth has resulted in the development of several Monte Carlo simulation programs that include \texttt{NuTauSim}~\cite{Alvarez-Muniz:2017mpk}, \texttt{NuPropEarth}~\cite{Garcia:2020jwr}, \texttt{TauRunner}~\cite{Safa:2019ege, Safa:2021ghs}, and the \texttt{nuPyProp}~\cite{NuSpaceSim:2021hgs,Patel:2021tbd} module of \texttt{nuSpaceSim}~\cite{Krizmanic:2020bdm}.
The propagation of neutrinos through the Earth can produce secondary particles, among them more taus, which yields a guaranteed tau neutrino flux~\cite{Soto:2021vdc}.
Tau neutrino propagation in Earth benefits from tau neutrino regeneration in the production and decay process $\nu_\tau\to\tau\to \nu_\tau$~\cite{Halzen:1998be}.
In both the regeneration modeling of the $\nu_\tau$ energy from $\tau\to \nu_\tau$ and for the energy distribution of the hadronic shower of a detected tau decay, the polarization of the $\tau$ has a potential impact.
While $\tau$'s produced in very high energy weak interactions are nearly 100\% polarized, it has been noted in ref.~\cite{Safa:2019ege} that $\tau$'s experience some depolarization as a consequence of electromagnetic energy loss after their production in Earth.
In this article, we quantify the depolarization of $\tau$'s as a consequence of electromagnetic energy loss.

Tau depolarization in electromagnetic interactions are dominated by scatterings in which the incoming tau loses a substantial fraction of its energy.
Bremsstrahlung is suppressed relative to $e^+e^-$ pair production and photonuclear tau interactions as they propagate through materials~\cite{Dutta:2000hh}. 
Electron-positron production that yields a change in tau energy of more than 10\% are rare.
For example, for $10^9$ GeV Earth-skimming tau neutrinos incident at $10^\circ$, only 0.02\% of the pair production tau scatterings have significant energy loss.
On the other hand, photonuclear interactions have more frequent scattering in which the final tau energies are less than 90\% of their initial energies.
Thus, we focus on tau photonuclear energy loss in this article.

We begin with an overview of tau spin polarization to define our notation.
Following the work of Hagiwara, Mawatari, and Yokoya~\cite{Hagiwara:2003di}, our review of tau polarization in weak interactions extended to high energies affirms that VHE taus are nearly 100\% polarized when emerging from tau neutrino charged-current interaction. 
In~\cref{sec:tau-depolEM}, we extend the evaluation of tau polarization to tau photonuclear scattering and track tau depolarization using \texttt{nuPyProp} and \texttt{TauRunner}.
Our results are shown in~\cref{sec:results}, followed by our conclusions.
Details of the leptonic currents for weak and electromagnetic interactions that go into the polarization calculation are in~\cref{app:leptoniccurrent}.
Analytic approximations to evaluate the impact of fully polarized and fully depolarized taus when they decay are included in~\cref{app:nuenergyfraction}.

\section{Overview of tau spin polarization vector}

We follow the work of Hagiwara et al.~\cite{Hagiwara:2003di} to set up the initial equations needed to calculate the depolarization effect in tau propagation through materials.
In this section, we first start with the spin polarization vector for an outgoing $\tau$  from a $\nu_\tau$ charged-current (CC) interaction or a $\tau$ electromagnetic interaction and its role in tau decay distributions.
Later in the section, we review the polarization effect for CC interactions of ultra-high energy $\nu_\tau$. 

\subsection{Tau spin polarization and decays}

The spin polarization three-vector of the outgoing tau, in its own rest frame, can be written as~\cite{Hagiwara:2003di}
\begin{eqnarray}
\nonumber
 \vec{s} &=& (s_x,s_y,s_z) \\
  &=& \frac{P}{2}(\sin\theta_P\cos\phi_P,\sin\theta_P\sin\phi_P,\cos\theta_P), 
   \label{eq:azimuthal_spin}
  \\  &=& \frac{P}{2}(\Lambda\sin\theta_P,0,\cos\theta_P)\,,
 \label{eq:spinvector}
\end{eqnarray}
where the spin direction is relative to the final state tau momentum direction in the lab frame, taken to be the $z$-axis.
In~Eq. (\ref{eq:azimuthal_spin}), $\theta_P$ and $\phi_P$ are the polar and azimuthal angle of the spin vector in the $\tau$ rest frame, and $P$ is the degree of polarization. 
The polarization vector lies in the scattering plane~\cite{Hagiwara:2003di}, thus $\phi_P = 0$ or $\pi$.
Therefore in~Eq. (\ref{eq:spinvector}), $\Lambda$ takes value $+1$ or $-1$ according to the azimuthal angle. 

In what follows, we define ${\cal P}_{z} \equiv 2 s_{z}$.
For a single scattering, we denote the polarization as ${\cal P}_{i,z}$ where $i = \nu, \tau$ for CC and EM scattering, respectively.
Later in this section we evaluate ${\cal P}_{\nu,z}$, and in~\cref{sec:tau-depolEM} we compute ${\cal P}_{\tau,z}$.
The net polarization at decay is denoted as ${\cal P}_z$.
In the massless tau limit for $\nu_\tau$ CC interactions, the produced $\tau\equiv\tau^-$ is fully polarized, i.e., it is left-handed (LH) and 
\begin{equation}
    \vec{s}\simeq -\frac{1}{2}(0,0,1)\,,
\end{equation}
so $s_{z} = -\frac{1}{2}$ and ${\cal P}_{\nu,z} = -1$. 
The same is expected for the massless limit of $\tau \rightarrow \tau$ EM scattering.  

The quantity ${\cal P}_{z}$ enters into the energy distribution of the $\nu_\tau$ from tau decay.
The differential decay distribution of the tau as a function of $z_\nu\equiv E_\nu/E_\tau$ can be written in the form
\begin{equation} \label{eq:dGammadx2}
    \frac{1}{\Gamma}\frac{d\Gamma(\tau\to \nu_\tau)}{dz_\nu}= \sum_i B_i\Bigl(g_0^i(z_\nu)+{\cal P}_{z}g_1^i(z_\nu) \Bigr)\,,
\end{equation}
where $g_0^i(z_\nu)$ and $g_1^i(z_\nu)$ depend on $i$ decay channels with branching fraction $B_i$. 
The full decay width can be modeled as the sum of $\nu_\tau$ plus $e\bar{\nu}_e$, $\mu\bar{\nu}_\mu$, $\pi$, $\rho$, $a_1$, and $4\pi$ final states~\cite{Pasquali:1998xf,Bhattacharya:2016jce}.
The functions $g_0^i$ and $g_1^i$ for purely leptonic decays are included in \cref{app:nuenergyfraction}.
We evaluate ${\cal P}_{z}$ for multiple electromagnetic scatterings of the tau in \cref{sec:tau-depolEM}.

In \texttt{TauRunner}, the sum over decay channels is used to generate $z_\nu$, while in \texttt{nuPyProp}, a single channel for purely leptonic tau decay with a unit branching fraction is used.  While not obvious, a numerical implementation of Breit-Wigner smearing of the four semileptonic decay channels following ref. \cite{Jadach:1993hs} added to the leptonic distributions leads to a distribution that nearly matches the  
purely leptonic decay distribution of the tau~\cite{Patel:2021tbd}. More details appear in appendix \ref{app:nuenergyfraction}.
Thus, it is not surprising that evaluations using \texttt{TauRunner} and \texttt{nuPyProp} quantitatively agree for the results shown in this article.

We note that for $\bar\tau\equiv\tau^+$ decays, the differential decay distribution is 
\begin{equation} \label{eq:dGammadx2bar}
    \frac{1}{\Gamma}\frac{d\Gamma(\bar\tau\to \bar\nu_\tau)}{dz_{\bar\nu}}= \sum_i B_i\Bigl(g_0^i(z_{\bar\nu})-{\cal P}_{z}g_1^i(z_{\bar\nu}) \Bigr)\,,
\end{equation}
for $z_{\bar\nu}\equiv E_{\bar\nu}/E_{\bar\tau}$. Therefore, CC production of 
a right-handed (RH) $\bar\tau$ will yield the same decay distribution in $z_{\bar\nu}$ as the $z_\nu$ decay distribution of LH $\tau$.

\subsection{Scattering kinematics}

We first define the kinematical variables for the CC and EM interactions.
For an incoming $\nu_\tau$ momentum for CC interaction or $\tau$ momentum for EM interaction ($k$), target nucleon momentum ($p$) and outgoing tau momentum ($k'$) in the laboratory frame, we write
\begin{eqnarray}
k^\mu &=& (E_i,0,0,p_i) \nonumber\\
p^\mu &=& (M,0,0,0)\\
k^{\prime\, \mu} &=& (E_\tau,p_\tau \sin\theta,0,p_\tau\cos\theta) \nonumber\,.
\end{eqnarray}

Here $E_i$ and $E_\tau$ are the incoming neutrino/tau and outgoing tau energies in the laboratory frame, respectively.
For $\nu_\tau$ CC interactions, $k^2=0$, while for $\tau$ EM interactions, $k^2=m_\tau^2$.
In both CC and EM scattering cases, $k'^2=m_\tau^2$. 
The Lorentz invariant variables, in terms of energy and angles in the lab frame where the target is at rest, are given by
\begin{eqnarray}
\nu &=& p\cdot q/M = E_i y = (E_i-E_\tau) \nonumber\\
\label{eq:xyq2}
Q^2&=&-q^2=-(k-k')^2 \\
&=&2(E_i E_\tau-p_i p_\tau\cos\theta)-k^2-k'^2\nonumber\\
x &=& Q^2/(2p\cdot q)\nonumber\,.
\end{eqnarray}

\begin{figure}[t]
    \centering
    \includegraphics[width=0.45\textwidth]{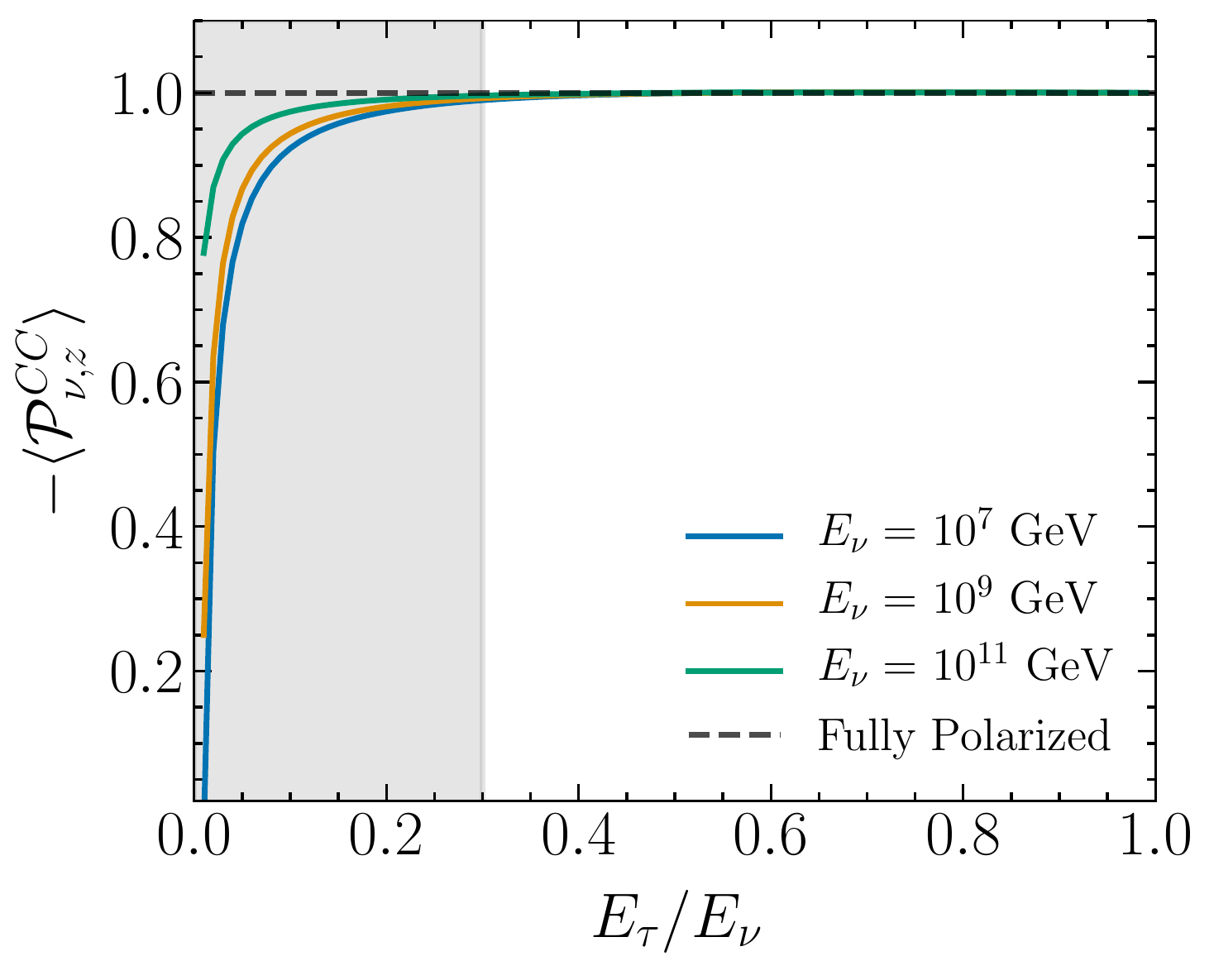}
    \caption{\textbf{\textit{Initial polarization of taus produced from charged-current scattering.}}
    The average value of $\langle{{\cal P}_{\nu,z}^{CC}}(y)\rangle$ as a function of tau energy fraction for $\nu_\tau$ CC scattering for different incident neutrino energies.
   The shaded region shows where 10\% or less of the taus emerge with this energy fraction.
    In the massless lepton limit, $-\langle {\cal P} _{\nu,z}^{CC}(y)\rangle = 1$ for all $y$,  corresponding to left-handed out-going taus.
    }
    \label{fig:neutrinopola}
\end{figure}

\subsection{Tau neutrino charged-current scattering} 
\label{subsec:neutrinoCC}

In this section, we review the calculation of the spin polarization vector components of the outgoing tau for $\nu_\tau N \rightarrow \tau X$ interaction~\cite{Hagiwara:2003di}, which will give us information on the polarization of the produced tau from an ultra-high energy $\nu_\tau$.
Polarization in neutrino CC production of taus has been discussed in refs.~\cite{Hagiwara:2003di,Kuzmin:2003ji,Kuzmin:2004ke,Kuzmin:2004yb,Graczyk:2004uy,Fatima:2020pvv}.
We perform the evaluation in the frame where the target nucleon is at rest. 
 
For neutrino charged-current interactions, in terms of lepton spinors, the leptonic weak current is
\begin{equation}
    j_\lambda^\mu(\nu\,, CC)= \bar{u}_\tau(k',\lambda)
    \gamma^\mu\frac{1-\gamma_5}{2}u_\nu(k,-)\,.
\end{equation}
Eqs.~\eqref{eq:neutplus} and~\eqref{eq:neutminus} give $j^{\mu}_\lambda$.
The leptonic tensor and hadronic tensor are expressed as  
\begin{eqnarray}
L_{\lambda\lambda'}^{\mu\nu}&=&j_\lambda^\mu j_{\lambda '}^{*\nu},\\
\nonumber
 W_{\mu\nu}&=&-g_{\mu\nu}W_1 + \frac{p_{\mu}p_{\nu}}{M^2}W_2 -\dot\iota\epsilon_{\mu\nu\alpha\beta}\frac{p^{\alpha}q^{\beta}}{2M^2}W_3 \\
 &+& \frac{q_{\mu}q_{\nu}}{M^2}W_4 + \frac{p_{\mu}q_{\nu}+q_{\mu}p_{\nu}}{2M^2}W_5\,.
\end{eqnarray}
The $\nu_\tau$ CC differential cross section in terms of inelasticity $y=(E_i-E_\tau)/E_i$ and $Q^2$ is 
\begin{equation}
    \frac{d^2\sigma_{\rm CC}}{dy\,dQ^2} = \frac{G_F^2}{4\pi}\Biggl(
    \frac{M_W^2}{Q^2+M_W^2}\Biggr)^2\frac{1}{E_\nu}F_\nu\,,
\end{equation}
where 
\begin{eqnarray}
\nonumber F_\nu &=& E_\nu\Bigl[(2W_1+\frac{m_\tau^2}{M^2}W_4)(E_\tau-p_\tau\cos\theta)\\
\nonumber
&+& W_2(E_\tau+p_\tau\cos\theta)\\
\nonumber
&+&\frac{W_3}{M}(E_\nu E_\tau+p_\tau^2-(E_\nu+E_\tau)p_\tau\cos\theta)\\
&-&\frac{m_\tau^2}{M}W_5\Bigr]\,.
\end{eqnarray}
Here, $F_\nu$ is obtained by contracting the hadronic and leptonic tensors with the appropriate normalization.
In what follows, we use the approximations $W_5=W_1$ and $W_4=0$ according to the Albright-Jarlskog \cite{Albright:1974ts} and Callan-Gross \cite{Callan:1969uq} relations which are exact in the massless parton, massless target approximations at leading order in QCD (see, e.g., ref. \cite{Kretzer:2002fr}).

The spin density matrix gives the relation~\cite{Hagiwara:2003di}
\begin{equation}
    dR_{\lambda\lambda'}\sim L_{\lambda\lambda'}^{\mu\nu}W_{\mu\nu}\,.
    \label{densitymatrix}
    \end{equation}
We can relate the elements of the spin polarization vector to the lepton-hadron contractions.
Up to an overall normalization factor $N$,
\begin{eqnarray}
dR_{++}+dR_{--} &=& d\sigma_{\rm sum} \label{densityA}\\ \nonumber
&=& N (L_{++}^{\mu\nu}+L_{--}^{\mu\nu})W_{\mu \nu},\\
dR_{+-} &=& s_x d\sigma_{\rm sum} \label{densityB}\\ \nonumber
&=& \frac{L_{+-}^{\mu \nu} W_{\mu\nu}}{(L_{++}^{\mu\nu}+L_{--}^{\mu\nu})W_{\mu\nu}} d\sigma_{\rm sum},\\
\frac{dR_{++}-dR_{--}}{2} &=& s_z d\sigma_{\rm sum} \label{densityC}\\
\nonumber
&=&  \frac{(L_{++}^{\mu\nu}-L_{--}^{\mu\nu})W_{\mu \nu}/2}
{(L_{++}^{\mu\nu}+L_{--}^{\mu\nu})W_{\mu\nu}}d\sigma_{\rm sum} \,.
\end{eqnarray}
Using the above equations, we can calculate the spin polarization vector components obtained for a single $\nu_\tau$ CC scattering
\begin{eqnarray}
\nonumber
s_x &=& -\frac{m_\tau}{2}\sin\theta E_\nu\Bigl[2W_1-W_2\\
&+&\frac{E_\nu}{M} W_3-\frac{m_\tau^2}{M^2}W_4+\frac{E_\tau}{M}W_5\Bigr]/F_\nu,\\
s_y &=& 0,\\ 
\nonumber
s_z &=& -\frac{E_\nu}{2}\Bigl[(2W_1-\frac{m_\tau^2}{M^2}W_4)(p_\tau - E_\tau\cos\theta)\\
\nonumber
&+& W_2(p_\tau+E_\tau\cos\theta)\\
\nonumber &+& \frac{ W_3}{M}((E_\nu+E_\tau)p_\tau - (E_\nu E_\tau+p_\tau^2)\cos\theta)\\
&-&\frac{m_\tau^2}{M}W_5\cos\theta\Bigr]/F_\nu\,.
\end{eqnarray}

One expects that the produced tau will be almost fully polarized for incident high-energy neutrinos~\cite{Payet:2008yu}.
In order to demonstrate this, we calculate the average value of ${\cal {P}}^{CC}_{\nu,z}$ as a function of $y$ by integrating over $Q^2$:
\begin{eqnarray}
\label{eq:calP}
\langle{{\cal {P}}^{CC}_{\nu,z}}(y)\rangle&\equiv &\frac{d\langle P\cos\theta_P\rangle_{\rm CC}}{dy}\\
\nonumber
&=& \int dQ^2 \, 2s_z \frac{d^2\sigma_{\rm CC}}{dy\, dQ^2}
\Biggl(\frac{d\sigma_{\rm CC}}{dy}\Biggr)^{-1}
\,.
\end{eqnarray}

\Cref{fig:neutrinopola} shows the average values of the $z$ component of spin polarization vector, $-\langle{{\cal P}_{\nu,z}}^{ CC}\rangle$, as a function of $E_\tau/E_\nu = (1-y)$.
As expected, it can be observed that the outgoing tau is almost fully polarized for high incident neutrino energies.
Depolarization occurs only for lowest energy fraction $E_\tau/E_\nu$, i.e., the largest $y$ values. 
The shaded band shows where 10\% or less of the $\tau$ emerge with this energy fraction.
Thus, it is a good approximation to assume $\nu_\tau\to\tau$ produces left-handed taus in agreement with results found in ref.~\cite{Payet:2008yu}.

The formulae presented here are specifically for $\nu_\tau\to \tau$, not $\bar\nu_\tau\to \bar\tau$. Antineutrino scattering involves a change of sign of the coefficient of the $W_3$ term. We consider here scattering with isoscalar targets, so the structure function $W_3$ depends only on the valence quark distributions. At high energies, valence contributions to the neutrino and antineutrino cross sections are small. For example, for incident neutrino and antineutrino energies of $10^7$ GeV, the CC cross sections differ by less than 1\% \cite{Gandhi:1998ri}. The results shown in ref. \cite{Payet:2008yu}, that the $\tau$ and $\bar\tau$ polarization magnitudes from $\nu_\tau$ and $\bar\nu_\tau$ CC interactions are equal at high energies, are therefore not surprising. For the remainder of the paper, we focus on the $\tau$ polarization with the understanding that the $\bar\tau$ polarization has the opposite sign.

\section{Tau depolarization in photonuclear scattering}
\label{sec:tau-depolEM}
As explained in the introduction, photonuclear interaction is the dominant electromagnetic energy loss mechanism for $\tau$'s for which depolarization effects can be observed.
Thus, in this paper we will only look at the photonuclear scattering process for taus. 

\subsection{Tau photonuclear scattering}

\begin{figure*}[t]
    \centering
    \includegraphics[width=0.9\textwidth]{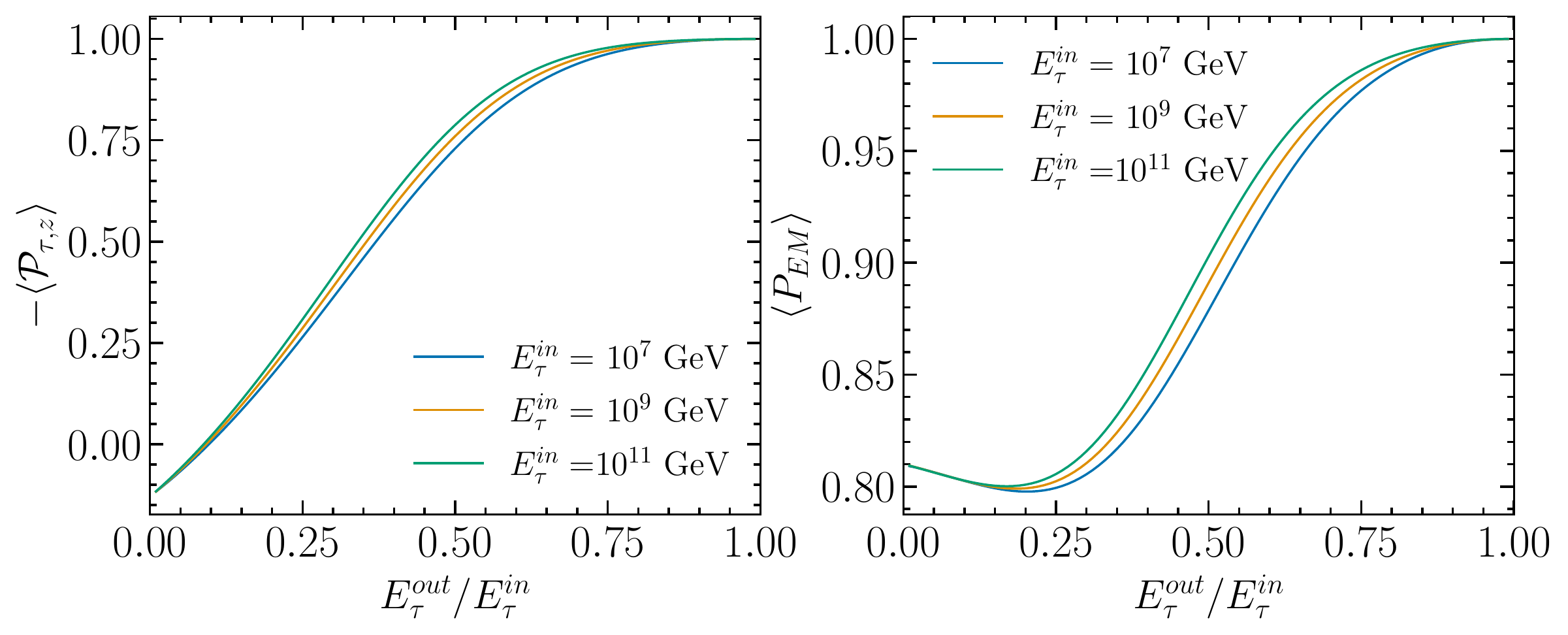}
    \caption{\textbf{\textit{Average polarization of outgoing tau after a single scattering.}}
    The average polarization about $z$-axis $\langle{{\cal P}_{\tau,z}}\rangle$ (left) and the average total polarization $\langle{P_{EM}}\rangle$ (right) as a function of outgoing tau energy fraction for single
    photonuclear electromagnetic scattering of $\tau$'s with rock for different incident tau energies. Here, $\langle{P^{EM}}\rangle=(\langle{{\cal P}^{EM}_{\tau,z}}\rangle^2$+$\langle{{\cal P}^{EM}_{\tau,x}}\rangle^2)^{\frac{1}{2}}$    
    }
    \label{fig:EMpol}
\end{figure*}
For tau electromagnetic scattering in the massless limit, again, the tau remains fully polarized if the initial state is polarized.
The inclusion of mass effects can diminish the magnitude of $P$ and introduce a $\theta_P$ dependence.
The quantities $P$ and $\cos\theta_P$ are functions of the outgoing tau energy and direction. 

The result obtained in the previous section for $\nu_\tau$ CC scattering gives us important information.
The first CC interaction of a high-energy cosmic neutrino passing through the Earth will produce a fully polarized $\tau$ and so it is safe to assume in tau EM photonuclear scattering the incoming tau is purely left-handed.
Following the same procedure as in~\cref{subsec:neutrinoCC} for $\nu_\tau\to \tau$, we derive the spin polarization vector from the spin density matrix for $\tau\to\tau$ electromagnetic scattering in terms of the structure functions. 

For tau electromagnetic interactions of initially left-polarized tau, the EM leptonic current is
\begin{equation}
    j_\lambda^\mu(\tau\,, EM)= \bar{u}_\tau(k',\lambda)
    \gamma^\mu u_\tau(k,-)\,
\end{equation}
The leptonic current expressions for EM interactions, $\lambda = \pm$, can be found in Appendix \ref{app:leptoniccurrent} in Eqs. (\ref{tauplus}), (\ref{tauminus}). 

The hadronic tensor for EM case will only have structure functions $W_1$, $W_2$ and reduces to
\begin{eqnarray}
 W_{\mu\nu}&=& -g_{\mu\nu}W_1+\frac{p_\mu p_\nu}{M^2}W_2,,
\end{eqnarray}
by taking into account gauge invariance. 

The differential cross section for tau photonuclear scattering  is 
\begin{equation}
    \frac{d^2\sigma_{\rm EM}}{dy\, dQ^2} = \frac{2\pi\alpha^2}{Q^4}\frac{1}{E_i}F_\tau\,,
\end{equation}
where
\begin{eqnarray}
\nonumber F_\tau &=& 2W_1(E_i E_\tau-p_ip_\tau\cos\theta -2 m_\tau^2)
\\ 
&+& W_2(E_i E_\tau + p_ip_\tau\cos\theta + m_\tau^2)\,,
\end{eqnarray}
is obtained by contracting the hadronic and leptonic tensors.
The structure functions $W_1$ and $W_2$ are more commonly written in terms of $F_1$ and $F_2$ as
\begin{eqnarray}
W_1 &=& F_1/M\\
W_2 &=& F_2/\nu \,,
\end{eqnarray}
with 
\begin{equation}
     F_1=\frac{1}{2x(1+R)}\Biggl(1+\frac{4M^2x^2}{Q^2}\Biggr) F_2\,.
     \label{eq:f1}
\end{equation}
The differential cross section translates to 
\begin{eqnarray}
\nonumber
    \frac{d^2\sigma_{\rm EM}}{dy\,dQ^2} &=& \frac{4\pi\alpha^2}{Q^4}\frac{F_2(x,Q^2)}{y}
    \Biggl[ 1-y-\frac{Q^2}{4E_i^2}
    \\ 
    &+& \Biggl( 1-\frac{2m_\tau^2}{Q^2}\Biggr)
    \frac{y^2(1+4M^2 x^2/Q^2)}{2[1+R(x,Q^2)]}\Biggr]\,,
\end{eqnarray}
where $x$, $y$, and $Q^2$ are related by~\cref{eq:xyq2}. The quantity $R(x,Q^2)$, implicitly defined
in eq. (\ref{eq:f1}), can be written terms of the longitudinal structure function $F_L\simeq F_2(x,Q^2)-2xF_1(s,Q^2)$ (in the small $x$ limit) and $F_1(x,Q^2)$. This
gives $R(x,Q^2)\equiv F_L(x,Q^2)/(2xF_1(x,Q^2))$.
At high energies, we can take $R(x,Q^2) = 0$.
Using Eqs. \eqref{eq:spinvector}, \eqref{densityA}, \eqref{densityB}, and \eqref{densityC} we can evaluate the spin polarization vector components for tau EM scattering case: 
\begin{eqnarray}
\nonumber
s_x &=& -\frac{m_\tau}{2}\sin\theta \Bigl[2E_i\, W_1
\\ \label{eq:sx}
&-&{W_2}(E_i +E_\tau)\Bigr]/F_\tau\\
s_y &=& 0 \\
\nonumber
s_z &=& -\frac{1}{2}\Bigl[2W_1(p_i p_\tau - E_i E_\tau\cos\theta)\\
\nonumber
&+& W_2(p_i p_\tau+E_i E_\tau\cos\theta)\\
\label{eq:sz}
&+& W_2 m_\tau^2\cos\theta \Bigr]/F_\tau\,.
\end{eqnarray}

The energy distribution of $\nu_\tau$ from $\tau$ decays depend on
${\cal {P}}_{\tau,z}$ (\cref{eq:dGammadx2}), so only the $s_z$ component of the spin polarization vector is relevant.
The average value of ${\cal {P}}^{EM}_{\tau,z}$ as a function of $y$, integrated over $Q^2$ is:
\begin{eqnarray}
\label{eq:calPEM}
\langle{{\cal {P}}^{EM}_{\tau,z}}(y)\rangle&\equiv &\frac{d\langle P\cos\theta_P\rangle_{\rm EM}}{dy}\\
\nonumber
&=& \int dQ^2 \, 2s_z \frac{d^2\sigma_{\rm EM}}{dy\, dQ^2} 
\Biggl(\frac{d\sigma_{\rm EM}}{dy}\Biggr)^{-1}\,
\,.
\end{eqnarray}

In~\cref{fig:EMpol} (left), the average value of $-\langle{{\cal P}_{\tau,z}^{EM}}(y)\rangle$ after a single scattering, as a function of $E_\tau^{out}/E_\tau^{in}$, is shown for three incident tau energies.
A high-energy tau passing through rock can get partially depolarized.
The depolarization effect becomes significant for $y \gtrsim 0.2$, namely, for $E_\tau^{out}/E_\tau^{in}\lesssim 0.8$. 

The value of $\langle {\cal P}_{\tau,z}^{EM}\rangle$ depends on the quantities $P$ and $\theta_P$. 
We calculate the average of the polarization, $\langle{P^{EM}}\rangle=(\langle{{\cal P}^{EM}_{\tau,z}}\rangle^2$+$\langle{{\cal P}^{EM}_{\tau,x}}\rangle^2)^{\frac{1}{2}}$, where $\langle{{\cal P}^{EM}_{\tau,x}}\rangle$ is evaluated according to~eq. (\ref{eq:calPEM}) with $s_z\to s_x$. 
In~\cref{fig:EMpol} (right), we show the average total polarization for different incident tau energies.
We see that  $\langle{P^{EM}}\rangle$ lies between 1 and 0.8 throughout the range of $E_\tau^{out}/E_\tau^{in}$.
The main contribution to depolarization observed in~\cref{fig:EMpol} (left) is from the polar angle $\theta_P$. 

\begin{figure*}[t]
    \centering
\begin{minipage}{0.45\textwidth}
\vspace*{-0.5cm}
    \includegraphics[width=0.95\textwidth]{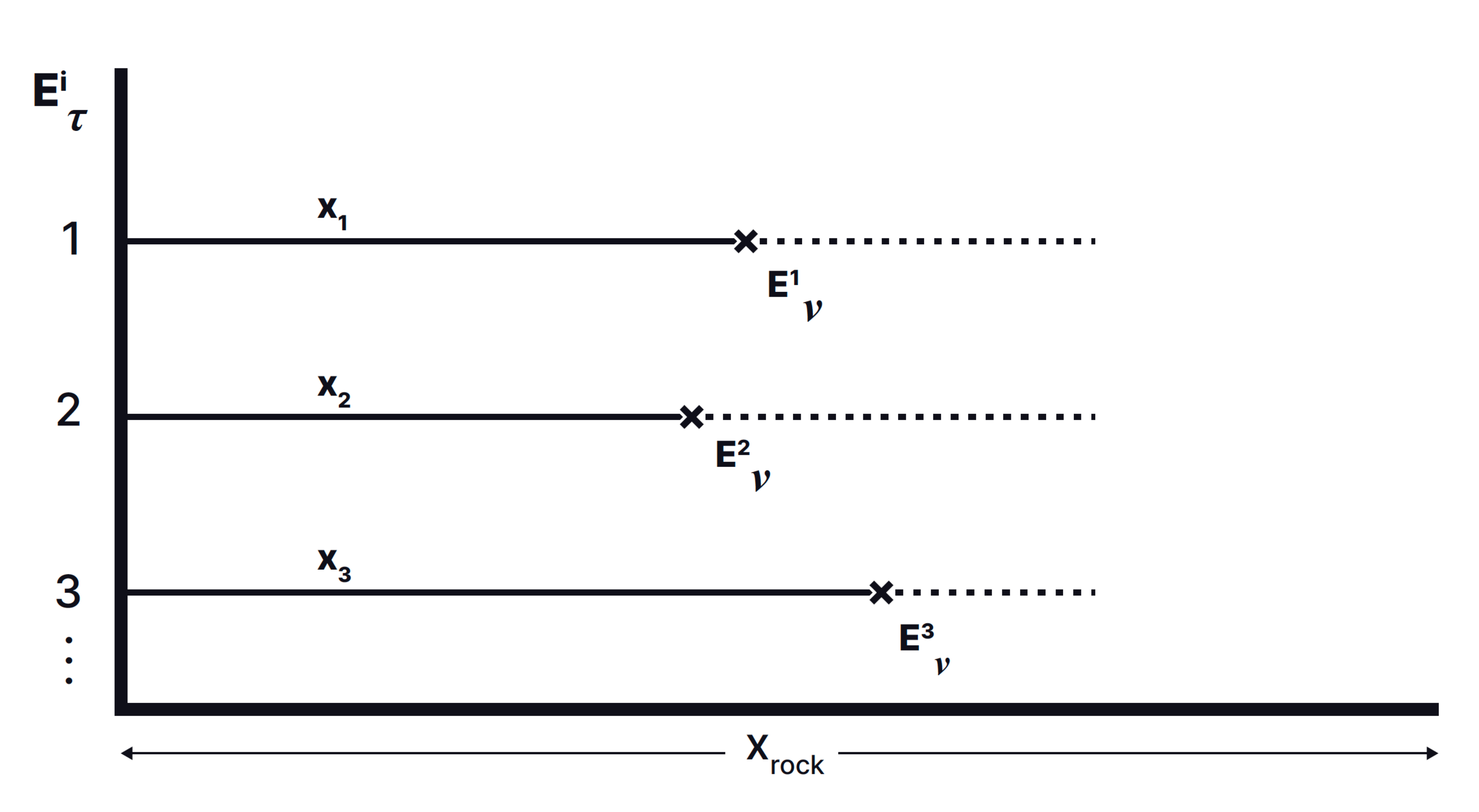}
    \end{minipage}
    \begin{minipage}{0.45\textwidth}
    \includegraphics[width=0.95\textwidth]{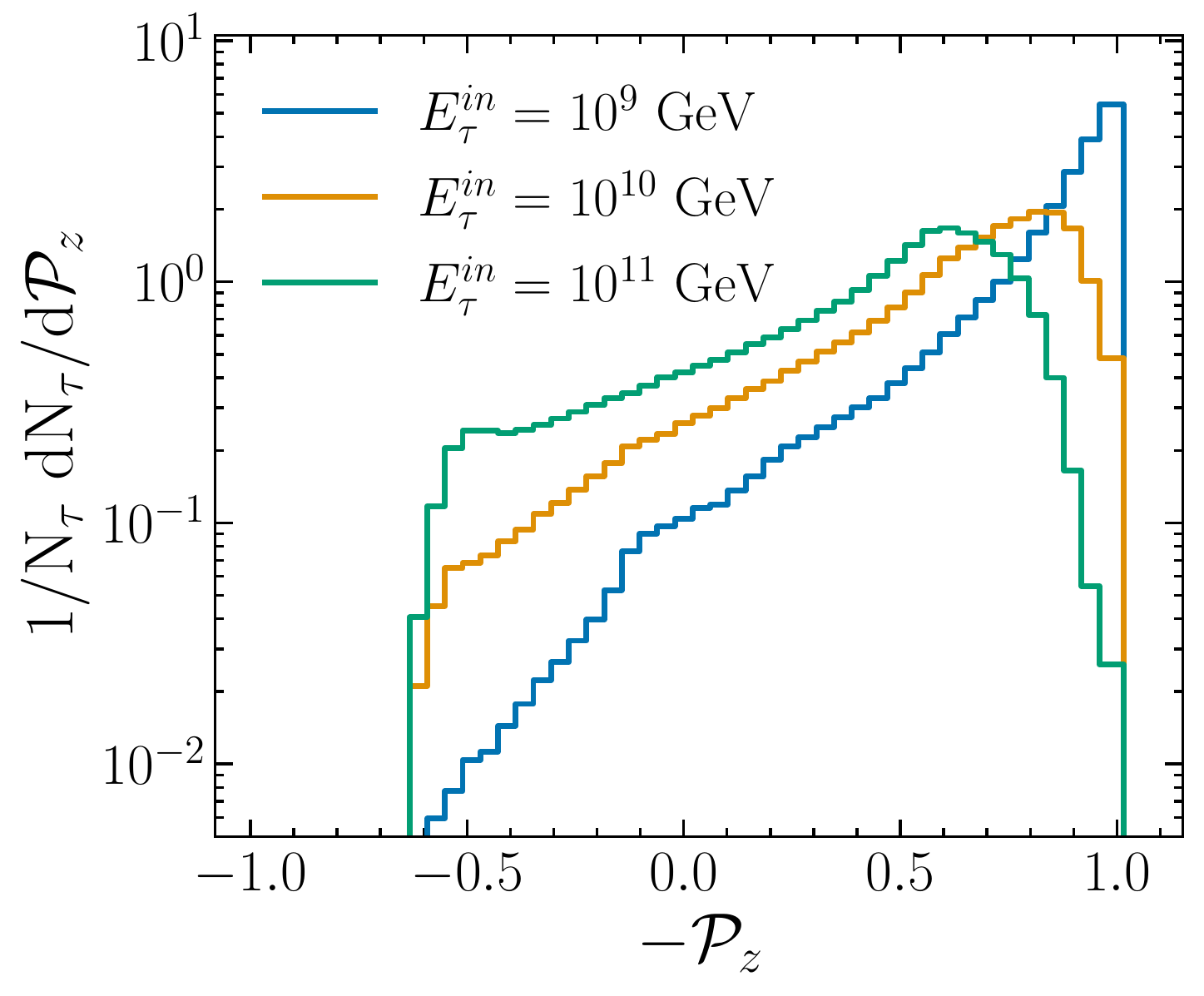}
    \end{minipage}
        \caption{\textbf{\textit{Schematic of taus entering a slab of rock and propagating until they decay, and the distribution of final polarization before tau decay.}} Left:
    Schematic of simulation of taus with $E_\tau^{in}$ incident on a thick slab of rock, in which they propagate varying distances, then decay. 
    Right: The differential number of taus as a function of final polarization from multiple EM interactions for $10^7$ $\tau$'s propagated through a slab of rock for each initial tau energy.}
    \label{fig:FinalPol}
\end{figure*}

\subsection{Monte Carlo implementation of tau depolarization}\label{sec:MC_depolarization}

There are multiple Monte Carlo packages that simulate propagation of the taus through the Earth.
Stochastic modeling is essential to incorporate the effects of tau depolarization since depolarization depends on $y=(E_\tau^{in}-E_\tau^{out})/E_\tau^{in}$.
Two such packages are \texttt{TauRunner}  and \texttt{nuPyProp}, which are modular PYTHON-based packages that track the propagation of charged leptons produced from charged-current interaction of neutrinos skimming through the Earth.
In this section, we focus on tau lepton propagation in rock, first to determine the distribution of the tau polarization just before it decays, then to illustrate impact of depolarization on the energy distribution of the tau neutrinos that come from tau decays.

We use \texttt{nuPyProp} and \texttt{TauRunner} to propagate $10^7$ tau leptons of each initial energy $E_{\tau}^{in}=10^9$, $10^{10}$ GeV and $10^{11}$~GeV through a slab of 200~km.w.e. of standard rock ($A=22$, $Z=11$, and $\rho=2.65\ {\rm g/cm^3}$), schematically illustrated in the left panel of~\cref{fig:FinalPol}.
With this depth of rock, all of the taus decay in the slab.
Tau propagation is performed  accounting for all electromagnetic energy loss processes.
As noted, for taus, $e^+e^-$ pair production and  photonuclear energy losses dominate, with photonuclear energy loss accounting for depolarization effects.
The simulation codes record $y$ (inelasticity) for each EM interaction of the tau.
For photonuclear interactions, the corresponding $\langle{{\cal {P}}^{EM}_{\tau,z}}\rangle$ and $\langle{P^{EM}}\rangle$ (see~\cref{fig:EMpol}) are used to determine $\theta_P$ and $P$ and with each photonuclear interaction, combined 
according to:
\begin{eqnarray}
   \label{eq:costhp} \cos\theta_{P} &=& \frac{\langle{{\cal {P}}^{EM}_{\tau,z}}\rangle}{\langle{P^{EM}}\rangle} \\
    \label{eq:thpf} \theta_{P,f} &=& \theta_{P,1} \pm \theta_{P,2} \pm \theta_{P,3} \pm ... \\
    \label{eq:pf} P_{f} &=& P_{1}\cdot P_{2}\cdot P_{3}\cdot ...\,,
\end{eqnarray}
where $\theta_{P,f}$ is calculated as the sum or difference of the polar angles because of the ambiguity in sign arising from cosine.
The sign is chosen using a random generator in the code.
We get the final polarization about $z$-axis for a single tau as
\begin{equation} 
\label{finalPolaeqn}
{\cal P}_{z} = P_{f}\cos\theta_{P,f}\,,
\end{equation}
where the tau had multiple interactions.
(For a single interaction of the tau, we have defined its polarization about $z$-axis with the notation ${\cal P}_{\tau,z}$.)
Using~\cref{finalPolaeqn} with our simulated data for $10^7$ incident taus, we show our results in the right panel of \cref{fig:FinalPol} for three incident tau energies, $E_{\tau}^{in}=10^9$~GeV, $10^{10}$~GeV, and  $10^{11}$~GeV.
It is observed that for $E_{\tau}^{in}=10^9$~GeV, the taus are more polarized and therefore the distribution of $-P_{z}$ peaks close to one.
On the other hand, for  $E_{\tau}^{in}=10^{11}$~GeV, there is more depolarization and we see a shift in the peak away from one.
This is due to the increase in number of interactions for higher initial tau energy which causes more depolarization. 

\begin{figure*}[ht!]
    \centering
    \includegraphics[width=0.8\textwidth]{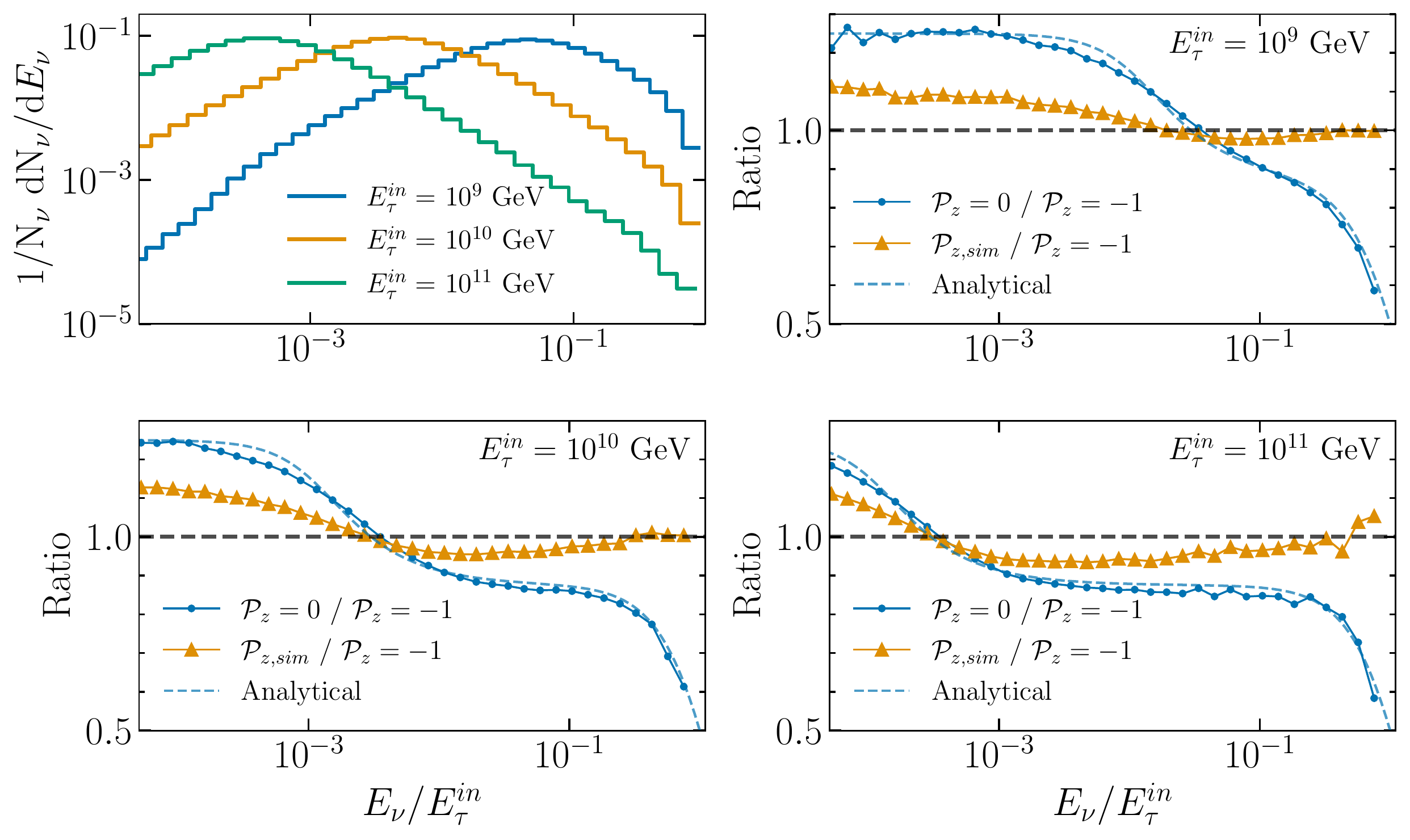}
    \caption{\textbf{\textit{Effect of tau depolarization on outgoing tau neutrinos for multiple energies.}}
    Upper left: Differential number of tau neutrinos as a function of $\nu_\tau$ energy fraction for simulated depolarization taus, shown for three different energies.
    Upper right: The orange (blue) markers show the ratio of the tau neutrino energy distribution with simulated (${\cal P}_z=0$) polarization to the tau neutrino energy distribution for left-handed tau decays as a function of $E_\nu/E_\tau^{in}$ for taus of energy $E_\tau^{in}=10^9$~GeV in rock.
    The dashed blue curve shows the approximate analytic evaluation of the ratio of decay neutrino distributions with ${\cal P}_z=0$ to ${\cal P}_z=-1$.
    The lower plots show these ratios for $E_\tau^{in}=10^{10}$~GeV (left) and $E_\tau^{in}=10^{11}$~GeV (right).}
    \label{Enu_plot}
\end{figure*}
The bump in the right panel of \cref{fig:FinalPol} for $E_{\tau}^{in}=10^{11}$~GeV at ${\cal P}_{z} \simeq -0.6$ arises because $\cos\theta_{P,f}$ can have negative values, i.e., $\theta_{P,f}>\pi /2$.
The distribution for $P_f$ is peaked at $\sim 0.6$ for $E_{\tau}^{in}=10^{11}$~GeV, which combined with negative values of $\cos\theta_{P,f}$ gives ${\cal P}_{z} \simeq -0.6$ for some fraction of taus.  

We turn to the neutrino energy distribution from the tau decays after propagating in rock.
For each incident tau at fixed energy, using the simulated data  that includes the final tau energy and polarization at the point of decay, we generate the energy of the tau neutrino from the tau decay.
This is done by creating a cumulative distribution function from the neutrino energy distribution equation given in~\cref{eq:dGammadx2}.

We show the effect of depolarization of taus on $\nu_\tau$ energy distribution in~\cref{Enu_plot}.
The upper left plot shows the differential number of tau-neutrinos as a function of tau-neutrino energy fraction ($E_\nu/E_{\tau}^{in}$).
Higher energy taus lose more of their initial energy as they propagate farther before their decays.

For three initial tau energies, the remaining  plots in \cref{Enu_plot} show ratio of $\nu_\tau$'s produced from unpolarized (${\cal P}_{z}=0$) to fully polarized (${\cal P}_z=-1$) taus (blue markers and curve), and simulated (${\cal P}_{z,sim}$) to fully polarized (${\cal P}_z=-1$) taus (orange markers and curve).
The cross-over points of the ratio plots approximately correspond to the peaks in the upper left plot of~\cref{Enu_plot}. 

To cross-check our results for the neutrino energy distribution from tau decays given taus incident on rock, we used an approximate analytical equation to get the $\nu_\tau$ spectrum.
Details are included in \cref{app:nuenergyfraction}.
Using~Eq. (\ref{eq:dNdZ}), we show dashed blue curves with this semianalytic approximation in \cref{Enu_plot}.
The ratio of the analytic evaluation of the neutrino energy distribution of unpolarized to fully left-handed polarized tau agrees very well with the ratio of distributions for ${\cal P}_z=0$ to ${\cal P}_z=-1$ in the Monte Carlo.

\begin{figure*}[t]
    \centering
    \includegraphics[width=0.35\textwidth ]{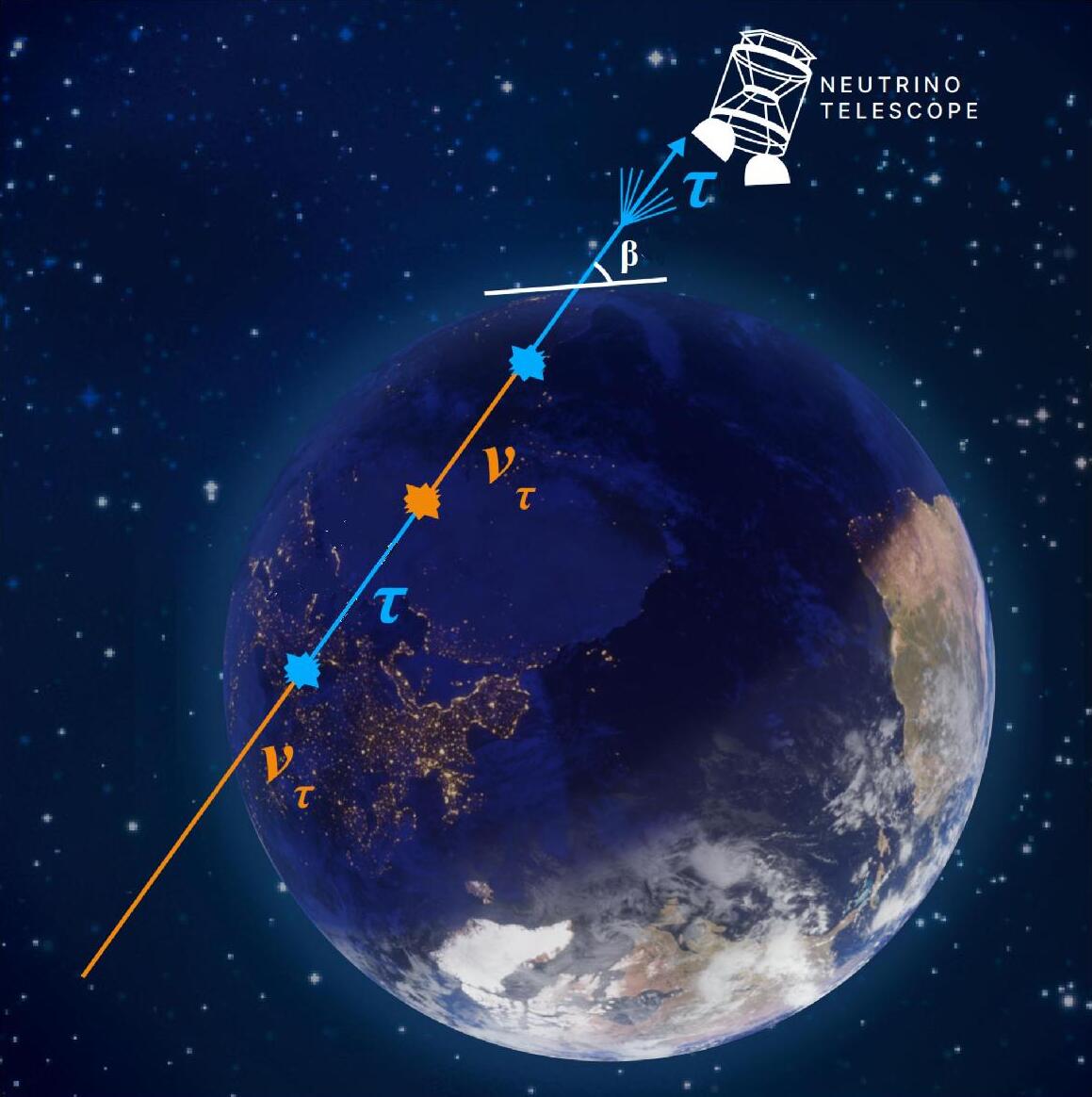} 
    \includegraphics[width=0.6\textwidth]{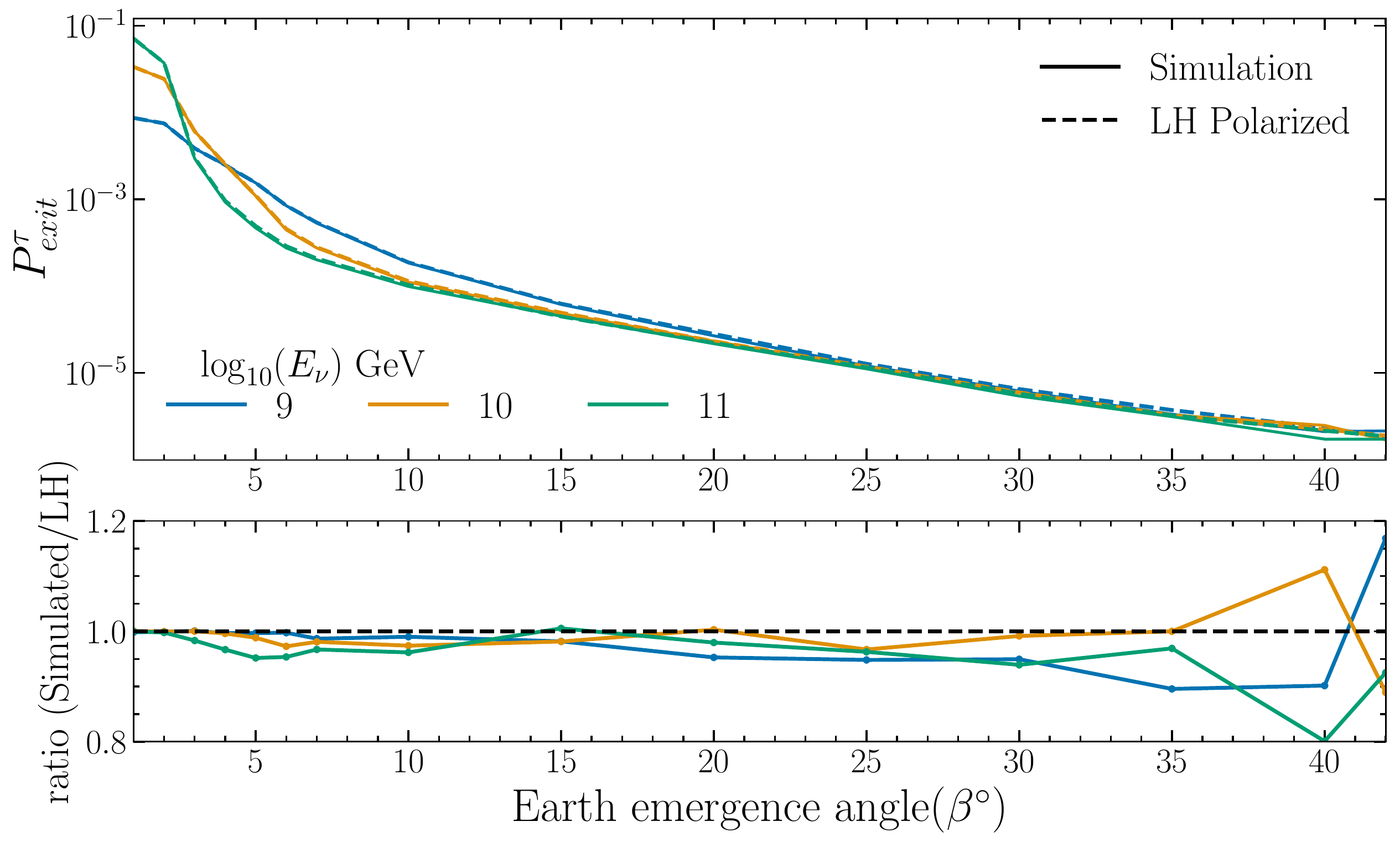}
    \caption{\textbf{\textit{Earth-skimming tau neutrino with emerging tau, and impact of tau depolarization on the tau exit probability.}}
   Left: Schematic diagram of a $\nu_\tau$ incident on the Earth that results in an emerging tau after series of CC interactions and tau decays (regeneration). The Earth emergence angle is $\beta$. 
   Right: Exit probability of taus as a function of different Earth emergence angles, for three different initial tau-neutrino energies. It shows a comparison when we consider LH polarization and simulated depolarization for EM interactions of the taus.}
    \label{fig:skimcartoon}
\end{figure*}

\section{Results for {Earth}-skimming tau neutrinos}
\label{sec:results}

\begin{figure*}[ht]
    \centering
    \includegraphics[width=0.45\textwidth]{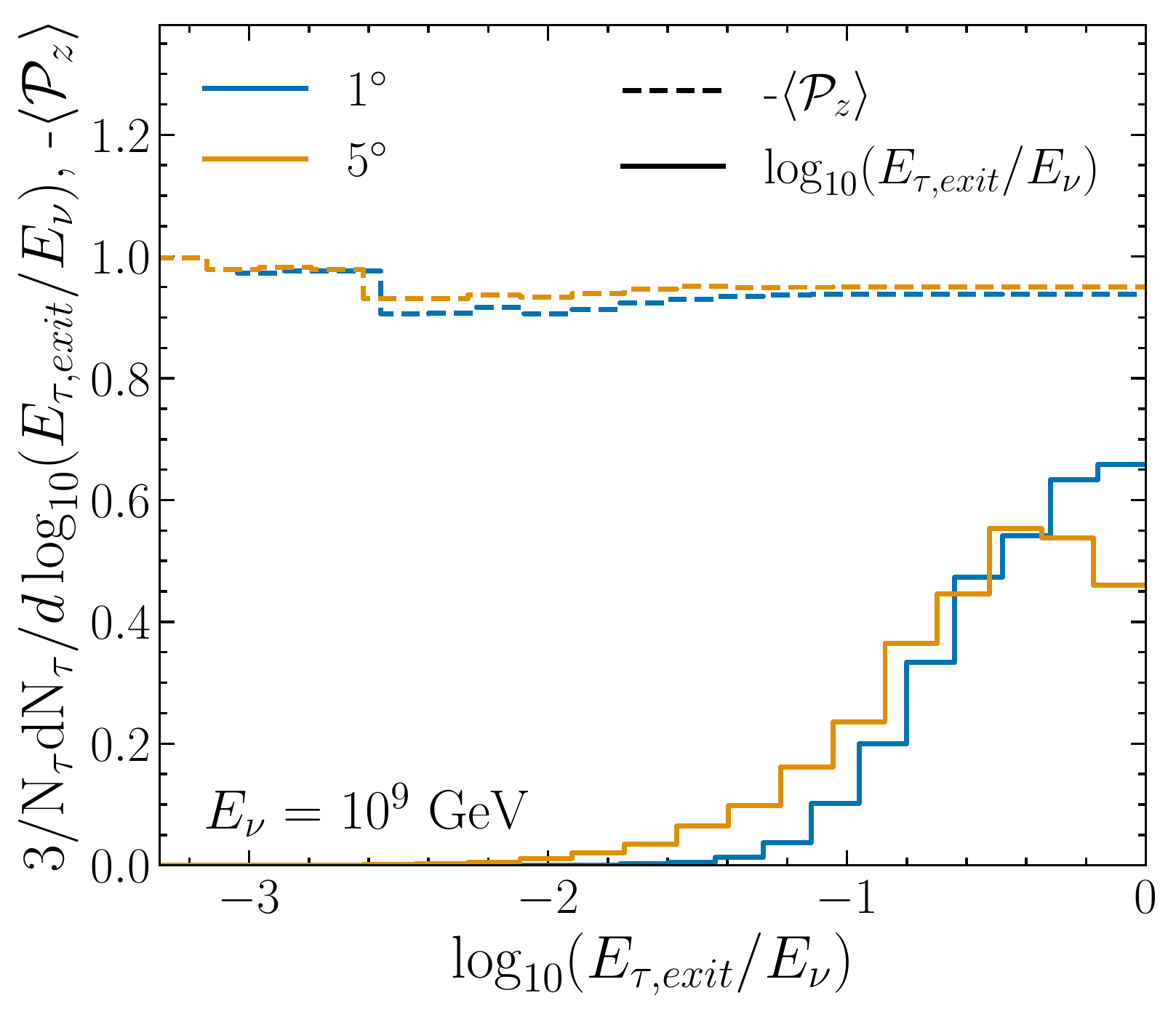}
    \includegraphics[width=0.45\textwidth]{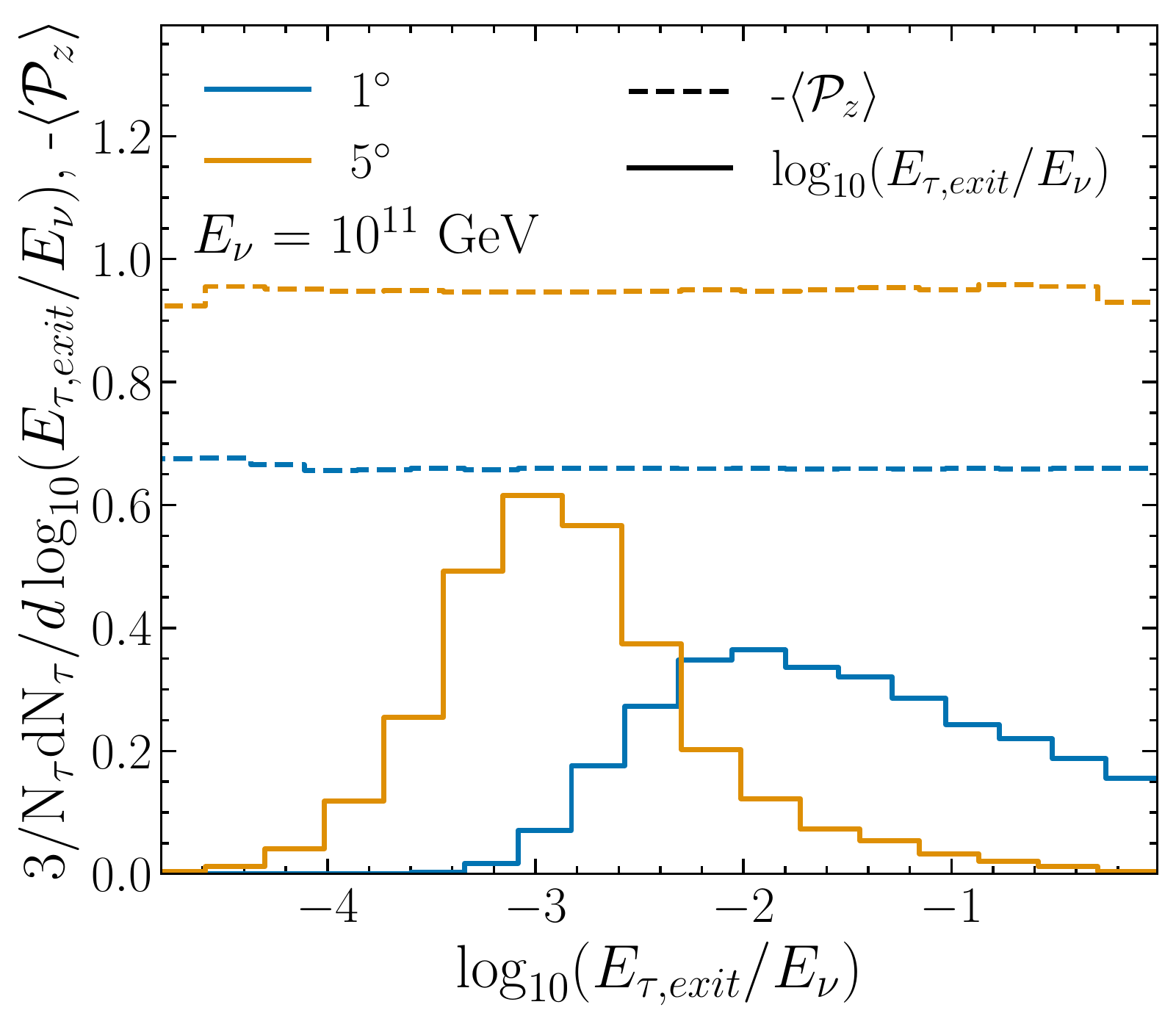}
    \caption{\textbf{\textit{Final energy of exiting taus and their polarization from a neutrino beam.}}
    Differential number of exiting taus, with a normalization of 3/N$_\tau$, and average polarization ($-\langle {\cal P}_z\rangle$) of exiting taus as a function of exiting tau's energy fraction for initial neutrino energies $E_\nu=10^9$~GeV (left) and $10^{11}$~GeV (right) and for $\beta=1^{\circ},5^{\circ}$.
    The energy distribution normalization was chosen so the energy distributions and polarizations can appear in the same figures. Note that the $x$-axes have different ranges in the two panels.}
    \label{Efexit}
\end{figure*}

The methodology of the depolarization calculations established in the previous sections can be easily implemented in the context of neutrino telescopes.
Again using \texttt{nuPyProp} and \texttt{TauRunner}, we simulate $\nu_\tau$'s skimming through the Earth, at different Earth emergence angles ($\beta$), which interacts to produce taus.
The taus have electromagnetic interactions and can experience some depolarization.
If the tau decays, its decay tau neutrino is propagated in the Monte Carlo simulation to determine if it interacts to produce a lower energy tau, so called ``regeneration.''
A schematic of Earth-skimming tau neutrino trajectories to produce a tau that emerges from the Earth is shown in the left panel of \cref{fig:skimcartoon}. 

For Earth-based, suborbital and satellite instruments that detect signals of tau decay-induced extensive air showers, modeling requires the probability that a neutrino produces an exiting tau, the energy of the emerging tau, and its final polarization upon exit.
This last quantity enters into modeling the energy of the hadronic final state in the tau decay.

One feature of regeneration is that whenever a regenerated tau neutrino interacts to produce a regenerated tau, to a good approximation, that tau will be fully polarized (LH) as we showed in \cref{subsec:neutrinoCC}.
We follow the same procedure as described in~\cref{sec:MC_depolarization} to calculate the depolarization effect, given by~\cref{finalPolaeqn}, now accounting for the variable density of the Earth along the particle trajectory. 

The amount of regeneration depends on the incident neutrino energy and angle of incidence (equal to the Earth emergence angle, $\beta$).
Higher neutrino energies correspond to shorter neutrino interaction lengths, allowing for tau production and decay earlier along the trajectory than for lower neutrino energies.
On the other hand, for small angles, the column depth is too short for regeneration to occur.
Below $\beta\lesssim 4^\circ$, regeneration is negligible~\cite{NuSpaceSim:2021hgs,Patel:2021tbd}. 
For $E_\nu$ of $10^9$~GeV, regeneration occurs for $\beta\gtrsim 10^\circ$, while for $E_\nu= 10^{11}$~GeV, regeneration occurs for $\beta\gtrsim 4^\circ$. 

In the right panel of \cref{fig:skimcartoon}, the exit probability of the taus is plotted for different Earth emergence angles, for three different initial tau neutrino energies.
It shows a comparison between LH polarized and polarization simulated for EM interactions of the taus.
We observe that the exit probability is changed by 5\% for smaller angles, and 10\% for higher angles, when we consider depolarization in the EM interactions.
This shows us that depolarization has a small impact on the exit probability of the taus. 

\Cref{Efexit} shows us the final energy of the exiting taus with corresponding average polarization.
The energy distributions are normalized by 3/N$_\tau$ so that they appear on the same scale as the polarization curves included in the figures. The plot on left is for $E_\nu=10^{9}$ GeV for $\beta=1^\circ$ and $5^\circ$. For this energy, the regeneration rate is negligible for both angles. The taus that exit the Earth are the ones created from the initial tau neutrinos that interact close to the surface of the Earth. Thus there is no significant depolarization and the final energy of the exiting taus is close to the initial tau-neutrino energy. In the plot on right in \cref{Efexit} for $E_\nu=10^{11}GeV$,  for $\beta=1^\circ$, the exiting taus are created from the initial tau neutrinos which interact farther from the surface of the Earth. The high energy taus that are produced are partially depolarized as they propagate and lose energy on their way to exit the Earth. For $\beta=5^\circ$, the taus able to exit the Earth are created from the regenerated tau neutrinos which interact closer to the surface of the Earth. Because of regeneration, the energy distribution of the emerging taus is lower than for $\beta=1^\circ$. Since the taus that emerge are produced close to the Earth surface, and with each tau production, its polarization is reset to $-1$, the average polarization for exiting taus is close to $-1$ for $\beta=5^\circ$ for incident neutrinos with $E_\nu=10^{11}$ GeV.

\section{Conclusions}

In principle, tau depolarization effects can affect the flux normalization and the energy distributions of tau neutrinos and taus that arrive at underground detectors or taus that emerge to produce up-going air showers.
We have performed an analysis of the dominant contribution to the depolarization of CC interaction produced left-handed taus as they transit materials. 

The depolarization of taus is not complete for tau energies up to $10^{11}$ GeV. With our simulations of tau energy loss in rock,
Fig. \ref{Enu_plot} shows that the neutrino energy distributions from tau decays are shifted from LH tau decays by at most $\sim \pm 10\%$.
The distortion of the neutrino energy distribution from the decays of fully polarized (LH) taus is less than the prediction for fully depolarized taus (${\cal P}_z=0$).
Even when taus are fully depolarized, the neutrino energy spectrum does not change significantly compared to the polarized distribution.

Taus that exit the Earth may come from a series of CC tau neutrino interactions and tau decays; this process is known as tau regeneration.
A consequence of the tau energy distribution from regeneration is an energy smearing that largely washes out the spectral distortion caused by the depolarization of taus.
We have shown that the tau exit probability is modified at most by $\sim 10\%$ for large Earth emergence angles, where the exit probability is already low. 
The energy distribution of the emerging taus is essentially the same with and without accounting for EM depolarization effects.

Our results for the polarization of the Earth-emerging taus show that the average polarization depends on the incident neutrino energy and angle, but it is largely independent of the final tau energy.
Improved modeling of the initial energy of the extensive air shower from tau decays in the atmosphere by including ${\cal P}_z$ is therefore straightforward to implement in Monte Carlo simulations with stochastic energy loss like \texttt{TauRunner} and \texttt{nuPyProp}. 

\acknowledgements
We thank Francis Halzen for useful conversations. We also thank Jackapan Pairin for producing the schematics in figure 3 and 5.
DG, SP and MHR are  supported  in  part  by  US  Department  of  Energy  grant DE-SC-0010113 and NASA grant 80NSSC19K0484.
CAA is supported by the Faculty of Arts and Sciences of Harvard University and the Alfred P. Sloan Foundation.
IS is supported by NSF under grants PLR-1600823 and PHY-1607644 and by the University of Wisconsin Research Council with funds granted by the Wisconsin Alumni Research Foundation.

\appendix
\section{Leptonic current for weak and electromagnetic scattering} \label{app:leptoniccurrent}

For completeness, we include the leptonic current for the weak interaction scattering $\nu_\tau N \rightarrow \tau^{-} X$  (eq. (20) in ref. \cite{Hagiwara:2003di}) 
\begin{eqnarray}
    \nonumber j_\lambda^\mu &=& \bar{u}_\tau(k',\lambda)
    \gamma^\mu\frac{1-\gamma_5}{2}u_\nu(k,-)\\
    j_{+}^\mu &=& \sqrt{2E_\nu(E_\tau - p_\tau)}\times \label{eq:neutplus}\\
\nonumber &&(\sin\thot,-\cos\thot,i\cos\thot,\sin\thot) \\
    j_{-}^\mu &=& \sqrt{2E_\nu(E_\tau + p_\tau)}\times \label{eq:neutminus}\\
\nonumber &&(\cos\thot,\sin\thot,-i\sin\thot,\cos\thot)
\end{eqnarray}
used to construct $L^{\mu\nu}_{\lambda \lambda'}$ for $\lambda,\lambda'=\pm$.

The
leptonic current for EM scattering $\tau^{-} N \rightarrow \tau^{-} X$ where the incident tau is left-handed ($\lambda'=-$), is
\begin{eqnarray}
\nonumber j_\lambda^\mu &=& \bar{u}_\tau(k',\lambda)
    \gamma^\mu u_\tau(k,-)\\
j_+^\mu &=& (\sqrt{\Sigma_i\Delta_f}-\sqrt{\Sigma_f\Delta_i})\times \label{tauplus}\\
\nonumber &&(f_+^0\sin\thot,-\cos\thot,i\cos\thot,\sin\thot)\\
j_-^\mu &=& (\sqrt{\Sigma_i\Sigma_f}-\sqrt{\Delta_i\Delta_f})\times \label{tauminus}\\
\nonumber &&(f_-^0\cos\thot,\sin\thot,-i\sin\thot,\cos\thot)
\end{eqnarray}
where
\begin{eqnarray}
\nonumber f_+^0 &=& (\sqrt{\Sigma_i\Delta_f}+\sqrt{\Sigma_f\Delta_i})/(\sqrt{\Sigma_i\Delta_f}-\sqrt{\Sigma_f\Delta_i})\\
\nonumber f_-^0 &=& (\sqrt{\Sigma_i\Sigma_f}+\sqrt{\Delta_i\Delta_f})/(\sqrt{\Sigma_i\Sigma_f}-\sqrt{\Delta_i\Delta_f})\ 
\end{eqnarray}
and with the definitions
\begin{equation}
\begin{matrix}
\Sigma_i=E_{i}+p_i & \Delta_i=E_{i}-p_i\cr
\Sigma_{f}=E_{\tau}+p_\tau & \Delta_f=E_\tau-p_\tau\,.
\end{matrix}
\end{equation}
With these definitions, $\sqrt{\Sigma_i\Delta_i}=m_\tau$. For neutrino scattering, $\Delta_i=0$ so eqs. (\ref{tauplus}) and (\ref{tauminus}) recover eqs. (\ref{eq:neutplus}) and (\ref{eq:neutminus}).

For high energies in electromagnetic scattering, $E_i >$ $10^7$ GeV, numerical cancellations are best handled by
making a Taylor expansion of the expressions up to order $O({m^2}/{E^2})$. In this approximation,
$s_z$ for $\tau$ EM scattering is 
\begin{eqnarray}
\nonumber
s_z &=& -\frac{1}{2 F_\tau}\Biggl[ W_1\Biggl(
Q^2-2Q_{\rm min}^2-2m_\tau^2\\
\nonumber
&+&\frac{m_\tau^2 Q^2}{2E_i^2}
\Bigl( 1+\frac{1}{(1-y)^2}\Bigr)\Biggr)\\ \nonumber
&+& W_2\Biggl(2E_i^2(1-y)\\
&-&\frac{Q^2}{2}\Bigl(1+
\frac{m_\tau^2}{2E_i^2}\Bigl(\frac{2-y}{1-y}\Bigr)^2\Bigr)\Biggr)\Biggr]
\end{eqnarray}
for $Q_{\rm min}^2=m_\tau^2 y^2/(1-y)$. We approximate $R=0$, so
\begin{eqnarray}
W_1 &=& \frac{E_iy}{Q^2}\Biggl( 1+\frac{Q^2}{y^2 E_i^2}\Biggr) F_2\\
W_2 &=& \frac{1}{E_i y} F_2\ ,
\end{eqnarray}
and
\begin{equation}
    W_1 = W_2 + \frac{E_i^2y^2}{Q^2} W_2\, .
\end{equation}

\section{{Tau survival and neutrino energies}}
\label{app:nuenergyfraction}

\begin{figure}[t]
    \centering
    \includegraphics[width=0.45\textwidth ]{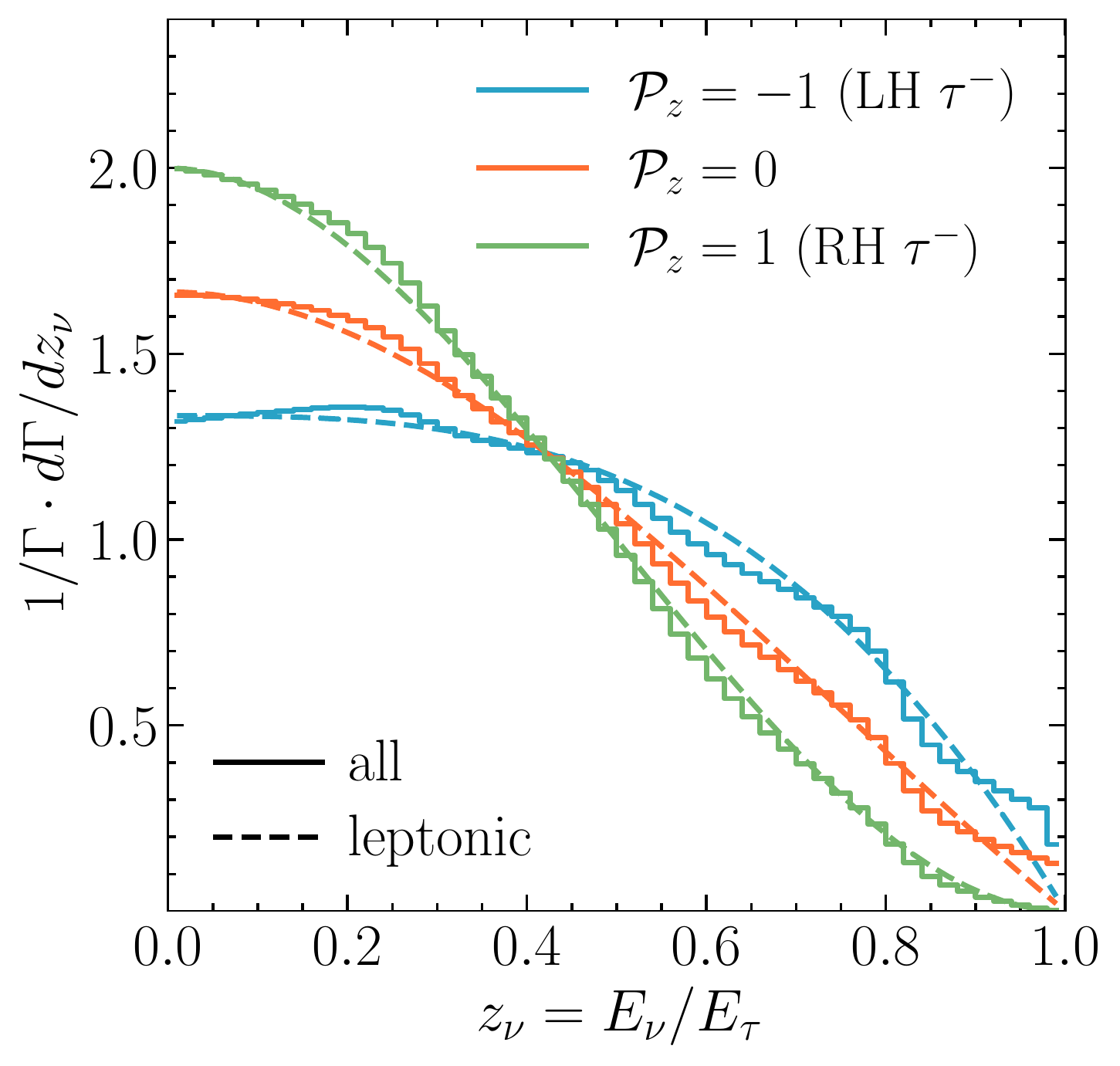} 
    \caption{\textbf{\textit{Tau decay distribution as a function of $z_\nu=E_\nu/E_\tau$.}} 
    The tau decay distribution as a function of $z_\nu$ for ${\cal P}_z=-1,\ 0,\ 1$ summed over all decay channels (solid line histograms, eq. (\ref{eq:dGammadx2})) \cite{Patel:2021tbd} and for the purely leptonic decay channels with unit branching fraction (dashed, eq. (\ref{eq:dGammadx}))
   }
    \label{fig:decaydist}
\end{figure}

As a cross-check to our Monte Carlo results, we have compared the tau neutrino energy distribution from polarized tau decays (${\cal P}_{z}=-1$) and unpolarized tau decays (${\cal P}_{z}=0$) from initially mono-energetic taus incident in rock with an approximate analytic evaluation. This approximate analytic expression accounts for tau energy loss according to 
\begin{equation}
\label{eq:dedx}
\Biggl\langle \frac{dE_\tau}{dX}\Biggr\rangle = -(\alpha+\beta E_\tau)\,,
\end{equation}
where $X$ is the column depth in g/cm$^2$.
We approximate the tau neutrino energy distribution with
\begin{equation} 
\label{eq:dGammadx}
    \frac{1}{\Gamma}\frac{d\Gamma(\tau\to \nu_\tau)}{dz_\nu}= \Bigl(g_0(z_\nu)+{\cal P}_{z}g_1(z_\nu) \Bigr)\,
\end{equation}
where $z_\nu=E_\nu/E_\tau$ and the functions $g_0(z_\nu)$ and $g_1(z_\nu)$
are the leptonic distributions \cite{Lipari:1993hd,Gaisser:2016uoy}
\begin{eqnarray}
g_0(z_\nu) &=& \frac{5}{3} - 3z_\nu^2+\frac{4}{3}z_\nu^3 \\
g_1(z_\nu) &=& \frac{1}{3} - 3z_\nu^2+\frac{8}{3}z_\nu^3\,.
\end{eqnarray}
The purely leptonic decay channels give a distribution that falls over the range from $z_\nu=0$ to $z_\nu=1$. For semileptonic decays to $\nu_\tau$ plus $\pi$, $\rho$, $a_1$ and $(4\pi)$, $1/\Gamma \cdot d\Gamma/dz_\nu$ ranges from $z_\nu=0$ to $z_\nu^{\rm max}$ where $z_\nu^{\rm max}=0.99,\ 0.81,\ 0.50$ and $0.29$, respectively \cite{Bhattacharya:2016jce}. In the absence of Breit-Wigner resonance smearing, this leads to $1/\Gamma\cdot d\Gamma/dz_\nu$ decreasing with steps at each $z_\nu^{\rm max}$ value. With Breit-Wigner smearing implemented as described in ref. \cite{Jadach:1993hs}, 
the purely leptonic $z_\nu$ distribution is a good approximation to the full neutrino distribution from the sum over all of the decay channels \cite{Patel:2021tbd}, as shown in fig. \ref{fig:decaydist}.

For high energy taus, we can 
neglect $\alpha$ and solve eq. (\ref{eq:dedx}) assuming that $\beta$ is energy independent to get 
\begin{eqnarray}
E_\tau(X)= E_{\tau}^{in}\exp(-\beta X)\,,
\label{eq:efx}
\end{eqnarray}
given $E_{\tau}^{in}$, the energy at $X=0$.
We approximate the energy loss parameter
$\beta=0.85\times 10^{-6}$ cm$^2$/g \cite{Dutta:2000hh}. The differential survival probability is
\begin{equation}
\frac{dP_{\rm surv}}{dE_\tau}\simeq  \frac{m_\tau}{c\tau\beta\rho E_\tau^2} P_{\rm surv}\,,
\end{equation}
which can be solved for constant $\beta$ to yield \cite{Dutta:2005yt}
\begin{equation}
P_{\rm surv}=\exp\Biggl[ -  \frac{m_\tau}{c\tau\beta\rho}\Biggl(
\frac{1}{E_\tau}-\frac{1}{E_{\tau}^{in}}\Biggr)\Biggr]\,.
\end{equation}
Thus, the differential survival tau probability is
\begin{equation}
     \frac{dP_{\rm surv}}{dE_\tau}=  \frac{m_\tau}{c\tau\beta\rho E_\tau^2}
    \exp\Biggl[ -  \frac{m_\tau}{c\tau\beta\rho}\Biggl(
    \frac{1}{E_\tau}-\frac{1}{E_{\tau}^{in}}\Biggr)\Biggr]\,.
\end{equation}

The differential number of tau neutrinos is 
\begin{eqnarray}
\nonumber 
\frac{dN_{\nu_{\tau}}}{dz}&=&E_{\tau}^{in}\int_{E_\nu/E_\tau^{in}}^1 \frac{dy_\nu}{z_\nu}\frac{1}{\Gamma}\frac{d\Gamma}{dz_\nu}\\ 
\label{eq:dNdZ}
&\times & \frac{dP_{\rm surv}}{dE_\tau} (E_\tau=E_\nu/z_\nu,E_\tau^{in})\,N_\tau(E_\tau^{in})\,.
\end{eqnarray}
Eq. (\ref{eq:dNdZ}) is used for the curves labeled ``Analytical'' in \cref{Enu_plot}.


\end{document}